# Mott insulating phase and coherent-incoherent crossover across magnetic phase transition in 2D antiferromagnetic CrSBr

F. Wu[1,2*], X. F. Zhang[1*], Y. Chen[1*], D. Pei[1,3], M. W. Zhan[1,4], Z.C. Tao[1], C. Chen[1], S. P. Lu[1,5,6], J. Z. Chen[7], S. J. Tang[5,6], X. Wang[8], Y. F. Guo[1], L. X. Yang[9], Y. Zhang[7], Y. L. Chen[1,10], Q. X. Mi[1†], G. Li[1†], Z. K. Liu[1†]

[1]*School of Physical Science and Technology, ShanghaiTech Laboratory for Topological Physics, ShanghaiTech University, Shanghai 201210, China*

[2]*Lingang Laboratory, Shanghai,200031, China*

[3]*Synchrotron SOLEIL, L'Orme des Merisiers, 91190, Saint-Aubin, France*

[4]*Shanghai Institute of Optics and Fine Mechanics, Chinese Academy of Sciences, Shanghai 201800, China*

[5]*State Key Laboratory of Materials for Integrated Circuits, Shanghai Institute of Microsystem and Information Technology, Chinese Academy of Sciences, 865 Changning Road, Shanghai 200050 China*

[6]*2020 X-Lab, Shanghai Institute of Microsystem and Information Technology, Chinese Academy of Sciences, Shanghai 200050, China.*

[7]*International Center for Quantum Materials, Peking University, Beijing, 100084, China*

[8]*Analytical Instrumentation Center, School of Physical Science and Technology, ShanghaiTech University, Shanghai 201210, China*

[9]*State Key Laboratory of Low Dimensional Quantum Physics, Department of Physics, Tsinghua University, Beijing, 100084, China*

[10]*Department of Physics, Clarendon Laboratory, University of Oxford, Parks Road, Oxford OX1 3PU, UK*

**In two-dimensional van der Waals magnetic materials, the interplay between magnetism and electron correlation can give rise to new ground states and lead to novel transport and optical properties. A fundamental question in these materials is how the electron correlation manifests and interacts with the magnetic orders. In this study, we demonstrate that the recently discovered 2D antiferromagnetic material, CrSBr, is a Mott insulator, through the combined use of resonant and temperature-dependent angle-resolved photoemission spectroscopy techiniques, supplemented by dynamical mean-field theory analysis. Intriguingly, we found that as the system transitions from the antiferromagnetic to the paramagnetic phases, its Mott bands undergo a reconfiguration, and a coherent-incoherent crossover, driven by the dissolution of the magnetic order. Our findings reveal a distinctive evolution of band structure associated with magnetic phase**

**transitions, shedding light on the investigation of intricate interplay between correlation and magnetic orders in strongly correlated van der Waals magnetic materials.**

**Introduction**

Strongly correlated electron systems exhibit a multitude of electronic phases, each characterized by unique physical properties such as high-temperature superconductivity [1], quantum spin liquid phase [2,3] and Mott transition [4,5]. The intricate interaction between charge and spin degrees of freedom sheds light on the phenomena of electron localization, representing one of the pivotal challenges in the study of strongly correlated materials. For instance, in cuprates, the ground state is shaped by the coexistence of magnetic order and strong electron correlations, which leads to a Mott insulating state [1,6,7]. Magnetism can influence the correlated electronic states in various ways, including gap formation [8], band folding [9,10], the creation of shadow bands due to magnetic scattering [11,12], and etc. A crucial aspect of Mott physics linked to magnetism is the emergence of local magnetic moments that remain coherent below the magnetic transition temperature ($T_c$) but fluctuate incoherently above it. This dynamic results in significant alterations in the electronic states mediated by magnetic scattering. Although theoretical models have extensively discussed these changes of electronic states in a Mott insulator due to magnetic transitions [13], direct experimental validation of these effects has yet remained scarce. This deficiency arises from the absence of an ideal magnetic compound that remains structually unchanged during the magnetic transition. Furthermore, the compound must be experimentally friendly, allowing for the access of its electronic structrure both above and below the magnetic transition.

The recently discovered two-dimensional (2D) van der Waals magnetic semiconductors provides new opportunities for exploring emergent quantum states, driven by the intertwined charge and spin degrees of freedom. These materials host a variety of phenomena, including low dimension magnetism [14,15], symmetry breaking [16] and excitonic insulator phase [17,18]. The pronounced electron interactions, amplified by the 2D quantum confinement, play a crucial role in establishing these distinctive quantum states. Notably, strong electron correlations have been evidenced in 2D magnetic semiconductors such as $CrGeTe_3$ and $CrI_3$, by both experimental and theoretical studies [19–22]. However, comprehensive investigations of the electronic structure alterations across magnetic phase transitions—which is vital for understanding the intricate interplay between magnetism and electron correlations—remains elusive.

Among all the van der Waals 2D magnetic semiconductors, CrSBr exhibits a relatively high magnetic transition temperature ($T_N \sim 132$ K, see Fig. S2(a,b)), a large magnetic moment (3 $\mu_B$/Cr) and a simple magnetic structure [23]. Previous optical measurements on this system have highlighted a strong magneto-optical coupling effect, indicating significant modulation of the electronic structure by the magnetic order [24–29]. Recent preliminary ARPES measurements have unveiled its general band structure, with a magnetic-field-induced splitted band observed 2 eV below the valence band maximum (VBM) [30]. However, the nature of correlation as well as the changes associated with the topmost valence band—primarily dominated by Cr $d$ orbitals remain unexplored experimentally.

In this study, employing both resonant and temperature-dependent ARPES measurements, alongside state-of-the-art DMFT calculation, we systematically explored the electronic band structure of CrSBr and its evolution across the antiferromagnetic (AFM) to paramagnetic (PM) phase transition. Resonant ARPES measurements suggest that the Cr-$t_{2g}$ predominantly form the topmost valence bands near the Fermi level ($E_F$), while the S/Br $p$ orbitals are more pronounced in the lower valence bands (E-$E_{VBM}$ = -2.5 eV ~ -1.5 eV, where $E_{VBM}$ is the energy position of the VBM). Temperature-dependent ARPES measurements reveal that the Cr-$t_{2g}$ bands undergo an additional energy shift (approximately 50 meV) as compared to that of the S/Br-$p$ bands during the magnetic phase transition. Moreover, the Cr-$t_{2g}$ bands exhibit broadening effects that exceed those induced by temperature alone when transitioning from the AFM to PM phases. These observations suggest a coherent-incoherent crossover of the Hubbard bands in CrSBr, a phenomenon corroborated by our DMFT calculations. Our findings not only underscore the unique coherent-incoherent crossover of Hubbard bands in CrSBr but also highlight its potential as a distinctive platform for study the interplay between electron correlation and magnetism.

**Methods**

*Crystal characterization:* The single crystals of CrSBr were grown by the chemical vapor transport (CVT) method (Details of the sample synthesis can be found in Supplementary Note S1). The sample typically appears as elongated strips [see Fig. S1(b)(i)] and its crystal structure was vertified by Bruker D8 single-crystalline x-ray diffraction (XRD) at ambient pressure. The X-ray diffraction patterns confirm the high quality of the crystals [Fig. S1(b)(ii-iv)]. The lattice constants were extracted as a = 0.351 nm, b = 0.477 nm, c = 0.8 nm, and α = 90°, β = 90°, γ = 90°, consistent with previous reports [23,31]. The magnetization measurement was conducted on a magnetic property measurement system (MPMS) and the specific heat measurement was performed on a physical property measurement system (PPMS-9).

*DMFT calculation:* The all-electron charge self-consistent DFT+DMFT calculations were performed on CrSBr with the embedded-DMFT package [32]. The DFT part for the DFT+DMFT calculations are prepared by the full-potential augmented plane-wave + local orbitals (APW+lo) program, WIEN2k [33]. We set RKmax = 7.0 and use a 20×15×4 k-mesh for Brillouin zone sampling. GGA-PBE scheme is used for approximating the exchange-correlation functional. In our DMFT calculations, we consider all five Cr-3$d$ orbitals with a density-density type local interaction between them. We set the on-site Coloumb interaction U = 4 eV and the Hund's exchange interaction J = 0.8 eV. The quantum impurity problem is solved by continuous-time quantum Monte Carlo (CT-QMC) method [34–37]. A 1x1x2 supercell, corresponding to the PM unit cell, were used for calculating PM and AFM phase on the same structure.

*ARPES experiment:* ARPES measurements were performed using DA30L electron analyzer with photon energies ranging from 40 to 60 eV at BL03U, SSRF and a helium discharging lamp with the photon energy of 21.2 eV at ShanghaiTech and Peking University. The samples were cleaved *in situ* and measured under ultrahigh vacuum with a base pressure below $6\times10^{-11}$ Torr. The energy and momentum resolution were 10 meV/7 meV/8 meV and 0.2°/0.2°/0.3° for the systems at BL03U/ShanghaiTech University/Peking University, respectively. To avoid the charging effect due to the semiconducting nature of CrSBr [38], we thinned down the sample to transparent flakes through mechanical exfoliation [see Fig. S1(b)(i)]. Photon-flux-dependent measurement was conducted to ensure the absence of the charging effect.

**Results**

The CrSBr crystallizes in an orthorhombic lattice with the space group of Pmmn (No.59) as shown in Fig. 1(a). The structure consists of layers stacked along the c-axis, and within each layer a Cr-S framework is sandwiched in between the planes of Br atoms. The magnetic properties are predominantly determined by the Cr atoms. At low temperatures, the magnetic moments within each layer are aligned coherently, while they are antiparallel between the layers, establishing an A-type AFM order [39], as depicted in Fig. 1(b).

The AFM order and strong electron correlations in CrSBr, which qualify it as a Mott insulator—a point that will be elaborated later—create a distinctive platform for exploring the interplay between the magnetic order and the strongly correlated electrons. This study highlights two major influences of the magnetic order on the correlated electronic band structure: 1) Magnetic order selectively enhances the energy shifts of the Hubbard bands (due to distinct orbitals) upon temperature increment. 2) Strong magnetic moment fluctuations above $T_N$ increases electron scattering, resulting in a coherent-incoherent crossover of Hubbard bands across the magnetic phase transition.

The magnetic order of CrSBr is characterized by measuring its temperature-dependent magnetic moments under a small external magnetic field of 0.1 T, applied along each crystallographic axis, as shown in Fig. 1(d). The AFM phase transition occurs at $T_N \sim 132$ K, consistent with previous reports [23,39]. The transition is further corroborated by specific heat measurements [Fig. S3(a)]. Density functional theory (DFT) calculations [Fig. 1(e)] suggest that CrSBr would be metallic in the PM state, with significant contributions from Cr $t_{2g}$ states near $E_F$. However, experimental efforts report a insulating behavior of the system with an optical bandgap of approximately 1.3 eV [31]. Incorporating electron interactions using dynamical mean-field theory (DMFT) leads to an insulating ground state [Fig. 1(e)], consistent with experimental result, therefore establishing the strongly correlated nature of its electronic states. The topmost valence bands (E-$E_F$ = -1.6 eV ~ -0.5 eV) is mainly contributed by electrons from Cr $t_{2g}$ orbitals, while the deeper valence bands (E-$E_F$ = -4 eV ~ -2 eV) mainly consists of electrons from $p$ orbitals from S and Br.

Figure 2(a) displays the ARPES spectrum of CrSBr in the AFM phase along the $\bar{X} - \bar{\Gamma} - \bar{X}$ direction, with the definition of the Brillouin zone (BZ) provided in Fig. S4(a). The VBM is located at the $\bar{\Gamma}$ point [zoom-in view is available in Fig. S4(d))] and the band gap size is estimated to be around 1.3 eV after the visualization of the conduction band with rubidium dosing [Fig. S4(e) and (f)]. The consistency between the ARPES results [Fig. 2(a)] and the DMFT calculation [Fig. 2(b)] allows for the identification of two primary groups of valence bands distinguishable with their orbital components: the first group consists of the Cr-$t_{2g}$ bands (labeled as γ) located from 0 to -1.1 eV relative to the VBM; and the second group consists of the S/Br-$p$ bands (labeled as α and β), ranging from -1.5 to -2.5 eV relative to the VBM.

The classification of these bands is further validated by resonant ARPES measurements, as shown in Fig. 2(c-f). When the energy of the incident photons was tuned to the resonance energy of Cr M-edge (3$p$ to 3$d$ transition, approximately 50 eV), a notable increasement of the photoemission intensity was evidenced on Cr-$t_{2g}$ bands [Fig. 2(c) and (d)]. Energy distribution curves (EDCs) at the $\bar{X}$ point demonstrate a 15 times enhancement of the γ state as compared to that of α and β states, indicative of its Cr-$d$ orbital origin [Fig. 2(e) and (f)].

To investigate the influence of magnetism on the band structure, temperature-dependent measurement was conducted across the AFM to PM phase transition, and the result is presented in Fig. 3. Figure 3(a-h) display the photoemission intensity along the $\bar{X} - \bar{\Gamma} - \bar{X}$ direction at various temperatures ranging from 79 K to 184 K, respectively. Notably, no obvious band merging in PM phase was evidenced, indicating a minor role of exchange band splitting due to magnetism in the low temperature AFM state. The intensity plots of the EDCs at the $\bar{X}$ point across the γ/α/β bands at different temperatures reveal their detailed evolution, as shown in Fig. 3(i). We observe that all γ/α/β bands shift towards the Fermi level upon increasing the temperature, accompanied with spectral weight broadening. Peak positions and the full width at half maximum (FWHM) of these bands are extracted through fitting with Lorentzian peaks and the result is plotted as a function of temperature in Fig. 3(j-l). We find the shifting rate of these bands with temperature is faster below $T_N$ and slower above. Moreover, the relative energy spacing between γ and the center of the α+β bands [detailed in the Fig. S7] exhibits a ~ 50 meV jump across $T_N$, indicating an orbital-selective enhancement of energy shift across magnetic transition. Furthermore, the spectral linewidth of the quasiparticle peak of the γ band [detailed in Fig. 3(l)] exhibits a significant increment with increasing temperature, with a surge of broadening rate above $T_N$, suggesting the influence from the loss of magnetic order, which increases electron scattering and a reduces the lifetime of the quasiparticles.

The observed orbital-selective band evolution and increment of spectral linewidth underscore the influence of magnetic order on the electronic structure of CrSBr. To elucidate the origins of this electronic structure evolution, we conducted DMFT calculations of the spectral function in both the AFM and PM phases, as shown in Fig. 4(a-b). The band structure shows nice agreement with the measured ARPES spectrum. Notably, the γ band demonstrates a relative upward shift compared to the α/β bands, and its FWHM is significantly broader in the PM phase. The calculated positions and FWHMs of the γ band, depicted in Fig. 4(c), exhibit similar trends to those observed experimentally [See Fig. S8 for details on the analysis]. Further, the calculations of the imaginary part of the self-energies highlight the Mott nature of the bands from the $t_{2g}$ orbitals relative to other orbitals. In the PM phase, the imaginary part of the self-energies of the $t_{2g}$ orbitals displays singularities within the energy gap, indicating a significant influence from many-body interactions and confirming the Mott band nature of these orbitals [Fig. 4(d)]. Conversely, in the AFM phase, the imaginary part no longer diverges, suggesting an increase in quasi-particle lifetime and coherence, as illustrated in Fig. 4(e).

**Discussion**

The consistency between ARPES measurement and DFT + DMFT results, particularly the coherent-incoherent crossover, strongly suggest CrSBr to be a Mott insulator with enhanced local moment fluctuations above $T_N$. This conclusion is supported by following experimental observations: (1) the distribution of Cr $t_{2g}$ states observed in resonant ARPES measurement and DFT + DMFT calculations, (2) the orbital-selective enhancement of band energy shifts as a function of temperatures, and (3) the coherent-incoherent crossover observed in the electronic states.

In the absence of electronic correlations, CrSBr would exhibit metallic behavior, characterized by a pronounced density of states on $E_F$. However, the PM phase DFT + DMFT calculations predict that electronic correlations split the Cr-$t_{2g}$ bands, forming the lower Hubbard bands as the top valence bands, as confirmed by the resonant ARPES measurement. This is a hallmark of typical Mott behavior, where the gap size continuously evolves with changes in temperature. Figure 3(j-l) and Figure 4(c) clearly demonstrate that the Cr-dominated bands, specifically band $\gamma$, shift towards $E_F$ as temperature increases, aligning with the predictions of the Mott model. The Mott insulating character of CrSBr is further supported by the intriguing coherent-incoherent crossover. As electrons become localized, electronic correlations redistribute charge across different local orbitals to minimize the total energy and giving rise to local moments. Above $T_N$, these local moments fluctuate strongly, whereas below $T_N$, they behave coherently, forming long-range magnetic order. In CrSBr, particularly within the Cr-$t_{2g}$ orbitals, electronic states are scattered by these fluctuating moments, resulting in blurred spectra at higher temperatures, as shown in Fig. 3. However, the scenario alters below $T_N$: the local moments stabilize. Electrons then encounter a quasi-static local Zeeman field, akin to navigating a fixed magnetic landscape. This stabilization restores their single-particle nature, leading to sharper and clearer electronic states in ARPES. The enhanced resolution of the α/β band splitting below $T_N$ underscores the coherent scattering of electrons at low temperatures.

**Conclusion**

CrSBr stands out as an exemplary low-dimensional material exhibiting long-range magnetism. Our findings robustly support an interaction-driven Mott scenario. This long-range magnetism not only significantly alters the electronic states but also mediates them through coherent and incoherent scatterings. Our findings shed light on the intricate interplay between electronic correlations and band structure in CrSBr, offering new insights into Mott physics.

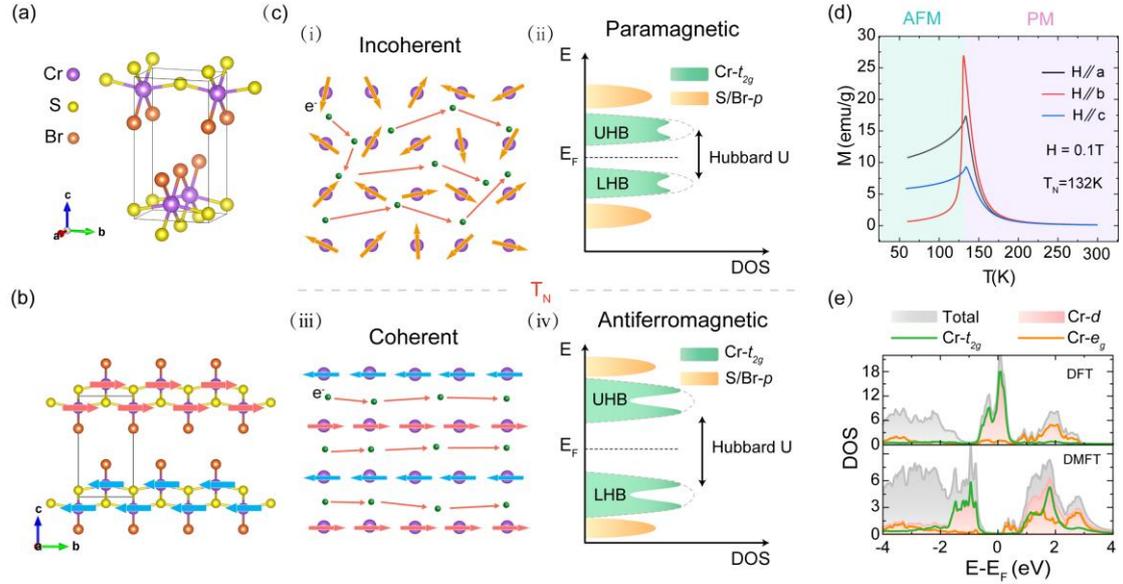

FIG.1. Concept illustration and physical properties of CrSBr. (a) Crystal structure of CrSBr. Purple, orange and yellow atoms stand for Cr, Br and S atoms, respectively. Black lines indicate the size of a unit cell. (b) Illustration of the anti-ferromagnetic structure of CrSBr in *b-c* plane. Red and blue arrows stand for the magnetic moments from Cr atoms. (c) Schematic of electrons scattering from fluctuating (i) and ordered (ii) magnetic momentum; as well as the depiction of density of state (DOS) in incoherent (ii) and coherent (iv) Mott states. (d) Plot of magnetic moment versus temperature under 0.1 T magnetic field along the *a*, *b* and *c* directions. (e) Calculated DOS using density function theory (DFT) and dynamical mean-field theory (DMFT). The regions of gray and pink are the DOS of total and Cr-*d* electrons, respectively, the green and orange curves are the DOS of Cr-$t_{2g}$ and Cr-$e_g$ electrons, respectively.

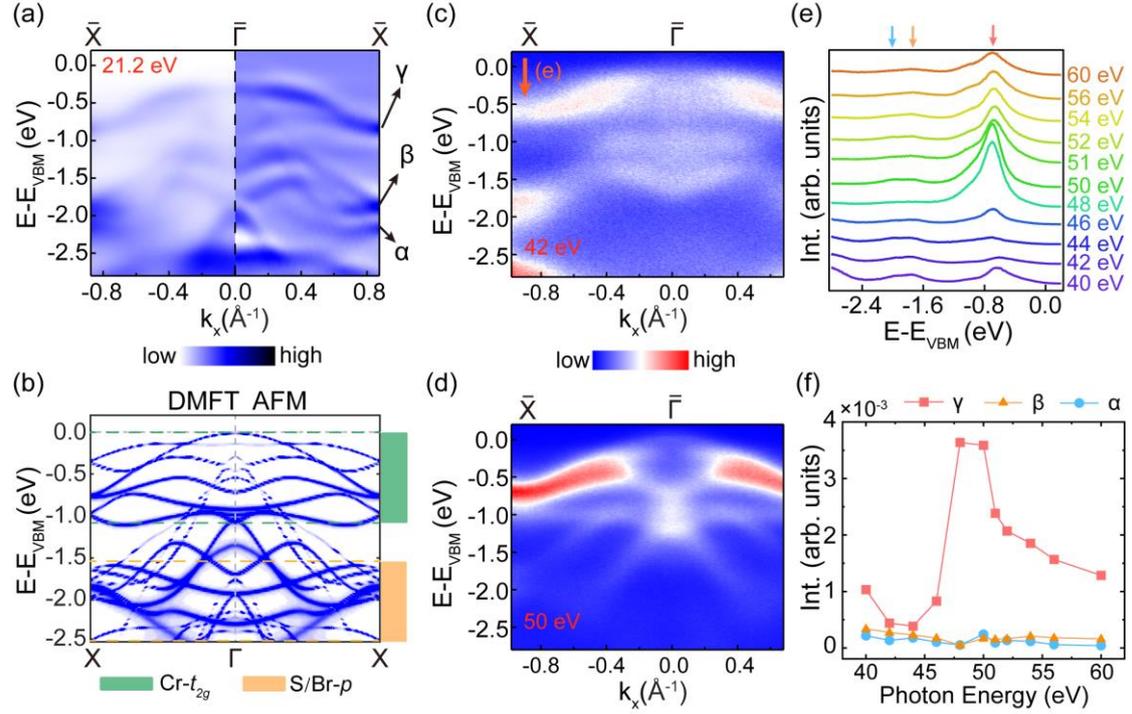

FIG.2. Electronic structure of CrSBr. (a) Photoemission intensity and its second derivative of the band structure along the $\bar{X}-\bar{\Gamma}-\bar{X}$ direction in CrSBr measured at 80 K. (b) Spectral function of CrSBr in antiferromagnetic phase calculated by DMFT at 93K. The areas of green and orange are mainly contributed by Cr-$t_{2g}$ and $p$ orbital originated from S and Br, respectively. (c) Off-resonant spectra along the $\bar{X}-\bar{\Gamma}-\bar{X}$ direction measured with 42 eV photons. (d) On-resonant spectra along the $\bar{X}-\bar{\Gamma}-\bar{X}$ direction measured with 50 eV photons. (e) The EDCs at $\bar{X}$ point measured with different photon energies. (f) Plot of the spectral weights of α, β, γ bands extracted from (e) as a function of photon energies. Data from resonant ARPES were collected at 90 K.

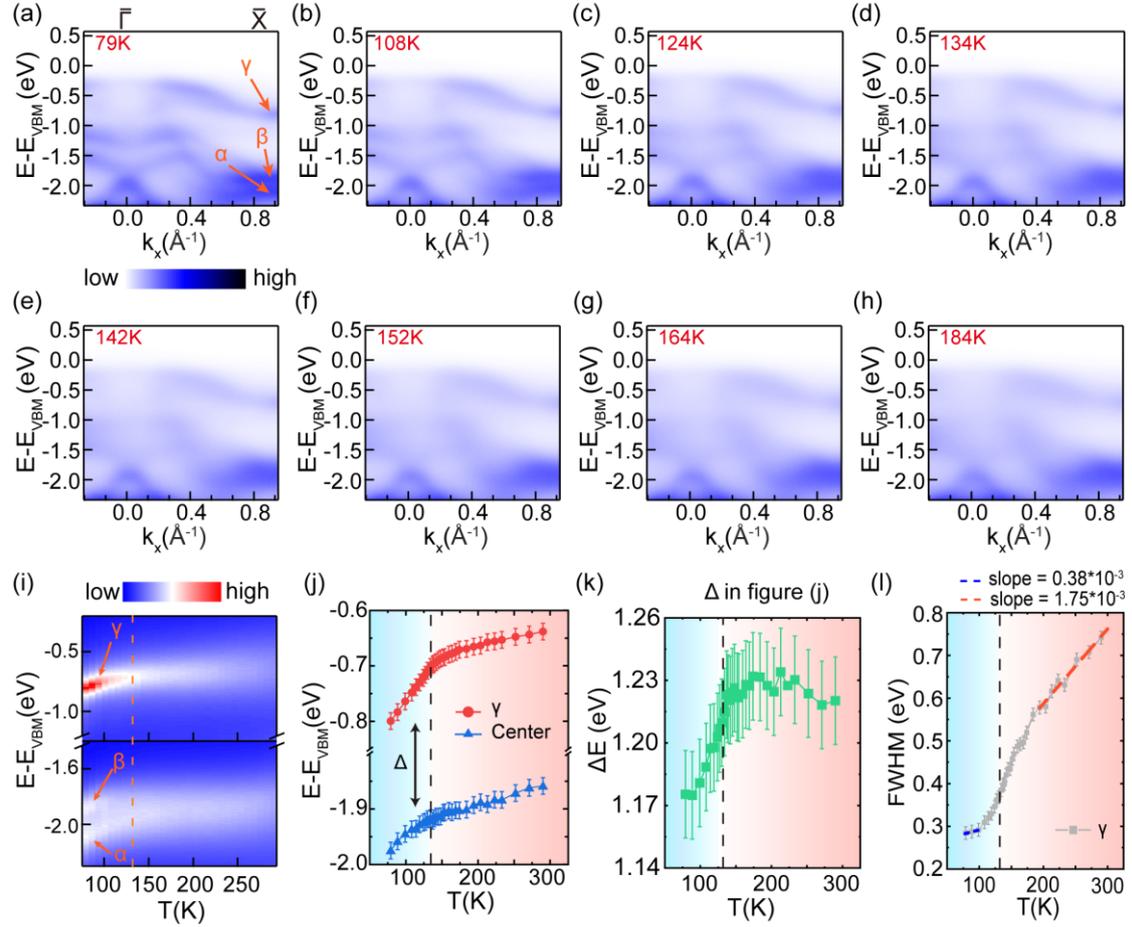

FIG.3. Temperature dependence of the band structure of CrSBr. (a-h) Band dispersion along $\bar{X} - \bar{\Gamma} - \bar{X}$ direction measured at T = 79 K, 108 K, 124 K, 134 K, 142 K, 152 K, 164 K, and 184 K, respectively. Bands α, β, γ are indicated by orange arrows in (a). (i) Intensity plot of the energy distribution curves for γ and α, β bands at $\bar{X}$ point as a function of temperature. (j) Temperature dependence plot of the extracted peak positions of the γ band, and the center position of the α and β bands. (k) Temperature dependence plot of the energy difference labelled in figure (j). (l) The gray curve stands for temperature-dependent FWHM, while the blue and orange dashed lines correspond to the linear fittings of data at low and high temperatures, respectively. Data shown were obtained using a He lamp (21.2 eV), with the valence band maximum (VBM) at 79 K set as the reference energy ($E_{VBM}$) for panels (a)-(j). Dashed lines in panels (i)-(l) corresponds to $T_N$=132K.

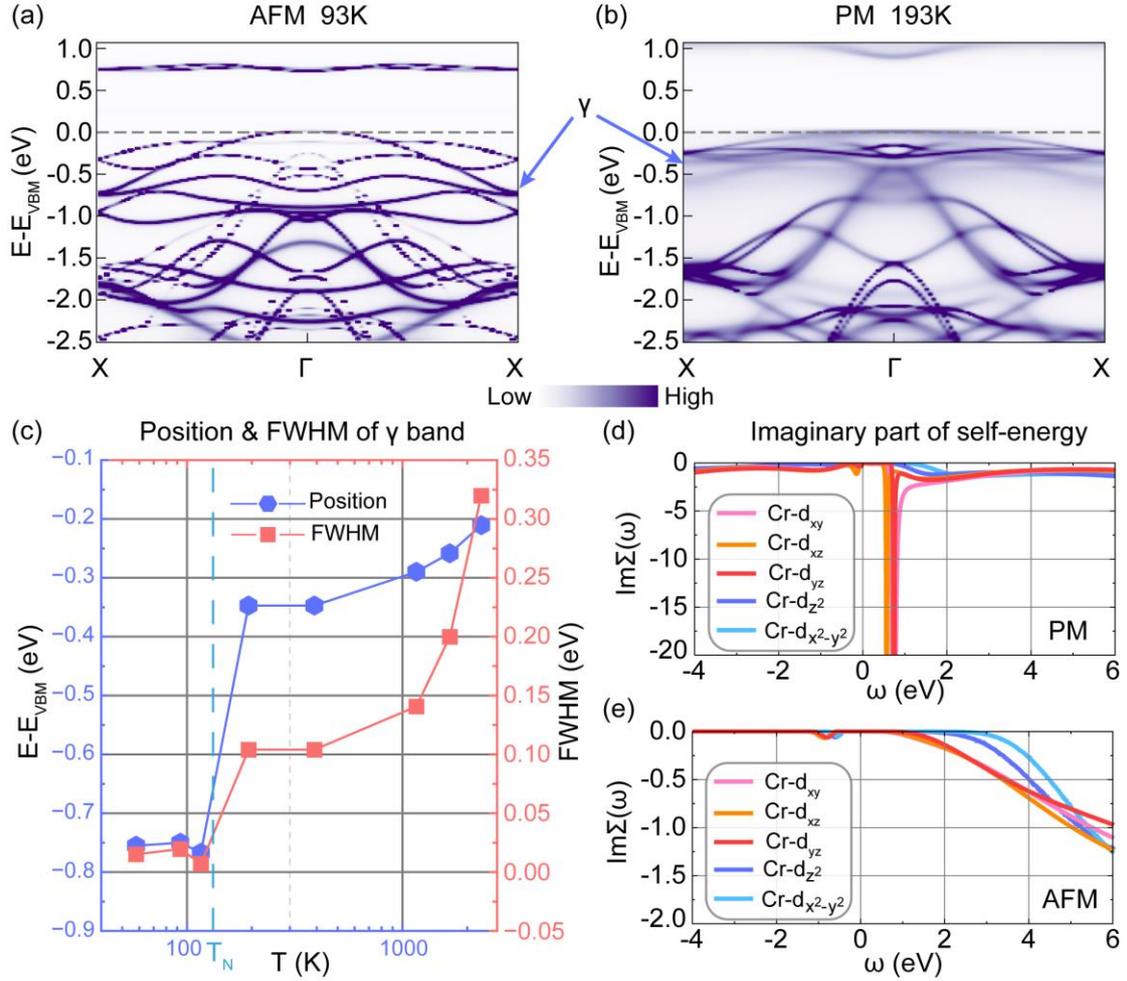

FIG.4. Calculated Electronic structure of CrSBr using density functional theory + dynamical mean-field theory (DFT+DMFT) under different temperature. (a) Spectral function of CrSBr in the antiferromagnetic phase at 93 K. Valence band maximum (VBM) is set as reference energy, and band γ is labeled with blue arrow. (b) Spectral function of CrSBr in the paramagnetic phase at 193 K. (c) Positions (blue hexagon) and full width at half maximum (FWHM, red square) of band γ at X point under different temperature. (d) Imaginary part of self-energy of Cr-$d$ orbitals calculated in the paramagnetic CrSBr at 193 K. (e) Imaginary part of self-energy of Cr-$d$ orbitals calculated in the antiferromagnetic CrSBr at 93 K. Only spin-up component is displayed for clarity.